\documentclass[aps,prd,floatfix,superscriptaddress,preprintnumbers,nofootinbib,twocolumn,longbibliography,dvipsnames,float,biblatex]{revtex4-2}
\pdfoutput=1

\usepackage{amsmath, amssymb, amsfonts, graphicx, mathrsfs, verbatim, float, slashed, bbm, xspace, enumerate, url, bbding, multirow, outlines, upgreek, braket,todonotes}
\usepackage{tabularx}
\usepackage{pgfplots}
\usepackage{tikz}
\usetikzlibrary{shapes.geometric, arrows,decorations.pathmorphing, decorations.pathreplacing, decorations.shapes, decorations.text, shapes.geometric, calc}
\tikzstyle{startstop} = [rectangle, rounded corners, minimum width=3cm, minimum height=1cm,text centered, draw=black, fill=orange!30, text width=5cm]
\tikzstyle{process} = [rectangle, rounded corners,minimum width=3cm, minimum height=1cm, text centered, draw=black, fill=blue!30, text width=5cm]
\tikzstyle{question} = [rectangle, rounded corners,trapezium left angle=70, trapezium right angle=110, minimum width=3cm, minimum height=1cm, text centered, draw=black, fill=red!30, text width=5cm]
\tikzstyle{arrow} = [thick,->,>=stealth]

\usepackage[]{xcolor} 
\usepackage[english]{babel}
\usepackage{hyperref}
\hypersetup{
     colorlinks   = true,
     citecolor    = blue,
     urlcolor     = blue,
     linkcolor    = blue
}
\newcolumntype{Y}{>{\centering\arraybackslash}X}

\begin{document}

\title{Deconstructive Composite Dark Matter Detection}

\author{Yilda Boukhtouchen}
\thanks{{\scriptsize Email}: \href{mailto:joseph.bramante@queensu.ca}{yilda.boukhtouchen@queensu.ca}; {\scriptsize ORCID}: \href{http://orcid.org/0009-0002-0414-1305}{0009-0002-0414-1305}}
\affiliation{The Arthur B. McDonald Canadian Astroparticle Physics Research Institute, \\ Department of Physics, Engineering Physics, and Astronomy,\\ Queen's University, Kingston, ON K7L 2S8, Canada}

\author{Joseph Bramante}
\thanks{{\scriptsize Email}: \href{mailto:joseph.bramante@queensu.ca}{joseph.bramante@queensu.ca}; {\scriptsize ORCID}: \href{http://orcid.org/0000-0001-8905-1960}{0000-0001-8905-1960}}
\affiliation{The Arthur B. McDonald Canadian Astroparticle Physics Research Institute, \\ Department of Physics, Engineering Physics, and Astronomy,\\ Queen's University, Kingston, ON K7L 2S8, Canada}
\affiliation{Perimeter Institute for Theoretical Physics, Waterloo, Ontario, N2L 2Y5, Canada}

\author{Christopher Cappiello}
\thanks{{\scriptsize Email}: \href{mailto:joseph.bramante@queensu.ca}{cappiellochristopher@gmail.com}; {\scriptsize ORCID}: \href{http://orcid.org/0000-0002-7466-9634}{0000-0002-7466-9634}}
\affiliation{Department of Physics and McDonnell Center for the Space Sciences, Washington University, St. Louis, MO 63130, USA}

\author{Melissa Diamond}
\thanks{{\scriptsize Email}: \href{mailto:joseph.bramante@queensu.ca}{melissa.d.diamond@mcgill@.ca}; {\scriptsize ORCID}: \href{http://orcid.org/0000-0003-1221-9475}{0000-0003-1221-9475}}
\affiliation{The Arthur B. McDonald Canadian Astroparticle Physics Research Institute, \\ Department of Physics\\ McGill University, Montreal, QC , Canada}

\begin{abstract}
We investigate the detection of composite dark matter that disassembles into a cascade while crossing the Earth. This occurs for loosely bound composite dark matter, where the binding energy per constituent is small, such that scattering with Standard Model nuclei typically imparts enough energy to dissociate a constituent from its composite. Trajectories and cascade profiles are found for dissociated constituents that are further diverted by scattering through the Earth. Such scattering cascades are a common feature of TeV-scale weakly-interacting dark matter loosely bound in composites. We identify underground detector signatures of constituent cascades that depend on composite characteristics; these signatures include  non-collinear multiple scatters in detectors, parameter-dependent timing separation of multiscatter events, and regions of parameter space where a dark matter cascade would leave a coincident signature in different underground laboratories.
\end{abstract}

\maketitle

\section{Introduction}

The nature of dark matter (DM), as well as its interactions, remain an enigma. Although many direct detection searches assume dark matter to be a point particle, dark matter that resides in composite states has been the subject of numerous studies \cite{Witten:1984rs,Farhi:1984qu,DeRujula:1984axn,Starkman:1990nj,Jacobs:2014yca,Krnjaic:2014xza,Detmold:2014qqa,Detmold:2014kba,Wise:2014ola,Wise:2014jva,Hardy:2014mqa,Hardy:2015boa, Gresham:2017cvl,Gresham:2017zqi,Bramante:2018qbc,Grabowska:2018lnd,Bramante:2018tos,Coskuner:2018are,Bai:2018dxf,Bramante:2019yss,Bai:2019ogh, Acevedo:2020avd,Cappiello:2020lbk,Acevedo:2021kly,Fedderke:2024hfy,Lu:2024xnb,Bleau:2025klr}.
Models consisting of asymmetric fermionic DM, bound together by a scalar field, have been primarily studied in the saturated composite regime, in which the binding energy per constituent $E_B$ is close to the mass of the individual fermionic constituents \cite{Wise:2014ola,Gresham:2017zqi}, making scattering off individual constituents suppressed \cite{Acevedo:2020avd,Acevedo:2021kly}. More recently, the interactions and formation of loosely bound composites, which have binding energies much smaller than the constituent masses, have begun to be explored \cite{Acevedo:2024lyr}. Compared to the often point-like interactions of tightly-bound composites, loosely bound composites have different interaction regimes, including an inelastic regime where constituents are scattered into excited states, and for sufficiently small constituent binding energies, a regime where constituents are ejected from the composite after scattering with Standard Model (SM) particles.

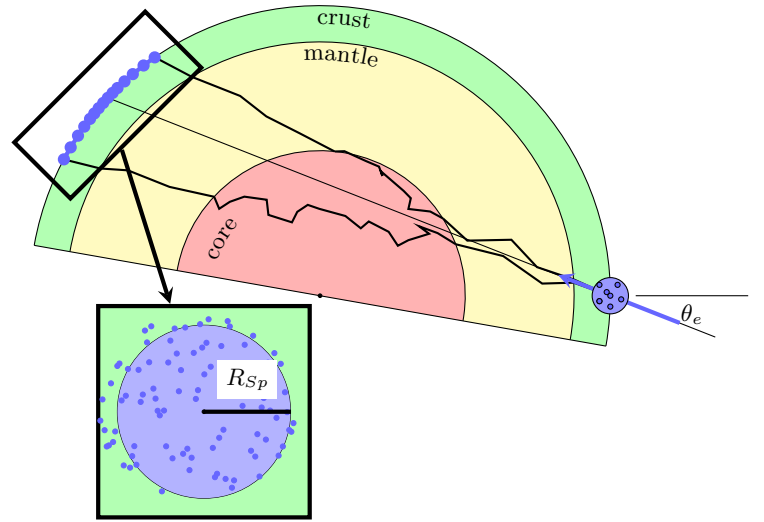
\begin{figure}[h!]
    \centering
    \def\MyScale{\columnwidth/\textwidth}
\begin{tikzpicture}[scale=\MyScale,
    declare function={a(\x)=sqrt(6.5^2 - \x^2);},
    declare function={b(\x)=-1*sqrt(6.5^2 - \x^2);}, 
    extended line/.style={shorten <=-#1},
  extended line/.default=1cm]
    \centering
    \coordinate (O) at (0,0);
    \draw[fill=green!30] (O) ++(170:8) arc (170:-10:8);
    \draw[fill=yellow!30] (O) ++(170:7) arc (170:-10:7);
    \draw[fill=red!30] (O) ++(170:4) arc (170:-10:4);
    \draw[fill=black] (O) circle (0.05);
    \draw (O) ++(170:8) -- (O);
    \draw (O) ++(-10:8) -- (O);
    \draw[decoration = {text along path, text align={align=center}, text={crust}}, decorate] (O) ++(100:7.5) arc (100:70:7.5);
    \draw[decoration = {text along path, text align={align=center}, text={mantle}}, decorate] (O) ++(100:6.5) arc (100:70:6.5);
    \draw[decoration = {text along path, reverse path, text align={align=center}, text={core}}, decorate] (O) ++(140:3) arc (140:160:3);

    \draw [extended line=1.5 cm] (8, 0) -- (-5.85, 5.46);
        % foreach \t in {0, 0.125, ..., 1} {
        % pic [pos=\t] {code={\draw circle [radius=2pt];}}
        % };
    ;
    \draw[thick, decorate,decoration={random steps,segment length=10pt,amplitude=0.1pt}] (8, 0) -- (6.99, 0.45) ;
    \draw[thick, decorate,decoration={random steps,segment length=15pt,amplitude=5pt}] (6.99, 0.45) -- (3.39, 2.13) ;
    \draw[thick, decorate,decoration={random steps,segment length=5pt,amplitude=5pt}] (3.39, 2.13) -- (0.4, 3.98) ;
    \draw[thick, decorate,decoration={random steps,segment length=15pt,amplitude=5pt}] (0.4, 3.98) -- (-3.54, 6.04) ;
    \draw[thick, decorate,decoration={random steps,segment length=10pt,amplitude=0.1pt}] (-3.54, 6.04) -- (-4.56, 6.58) ;

    \draw[thick, decorate,decoration={random steps,segment length=10pt,amplitude=0.1pt}] (8, 0) -- (6.99, 0.36) ;
    \draw[thick, decorate,decoration={random steps,segment length=15pt,amplitude=5pt}] (6.99, 0.36) -- (3.74, 1.42) ;
    \draw[thick, decorate,decoration={random steps,segment length=5pt,amplitude=5pt}] (3.74, 1.42) -- (-2.92, 2.73) ;
    \draw[thick, decorate,decoration={random steps,segment length=15pt,amplitude=5pt}] (-2.92, 2.73) -- (-6.06, 3.51) ;
    \draw[thick, decorate,decoration={random steps,segment length=10pt,amplitude=0.1pt}] (-6.06, 3.51) -- (-7.06, 3.76) ;

    \draw[blue!60, ultra thick] (O) ++(151.95:8) arc (151.95:124.73:8)
    foreach \t in {0, 0.1, 0.2, 0.28, 0.35, 0.4, 0.45, 0.5, 0.55, 0.6, 0.65, 0.72, 0.8, 0.9, 1} {
    pic [pos=\t] {code={\draw[fill=blue!60] circle [radius=1.5pt];}}
  };

    \node (zoom) at (-5.85, 5.26) [rectangle, minimum width=70, minimum height=30, draw, ultra thick, rotate=45] {};

    \node (zoomed) at (-3.2, -3.2) [minimum size=80, draw, ultra thick, fill=green!30] {};

    \draw[arrow, ultra thick] (zoom) -- (zoomed);

  \begin{scope}
  \clip [] (-3.2, -3.2) circle [radius=2.4];
  \draw[fill=blue!30] (-3.2, -3.2) circle (2.4);

  \end{scope}
  \draw [fill=black] (-3.2, -3.2) circle (0.05);

  \begin{axis}[
    domain=-7:7,
    anchor=target,
    axis lines=none,
    axis equal image,
    xtick=\empty, ytick=\empty,
    clip mode=individual, clip=false
]
% \addplot [thick] {a(x)};
% \addplot [thick] {b(x)};
\addplot [blue!60, only marks, mark=*, samples=100, mark size=2]
    {0.5*(a(x)+b(x)) + 0.5*rand*(a(x)-b(x))};
\node[coordinate] (target) at (axis cs:1.3*\MyScale,2*\MyScale){};
    \end{axis}

  \draw[ultra thick] (-3.2, -3.2) ++(0:2.4) -- node[anchor=south, fill=white!90, outer sep = 5] {$R_{Sp}$} (-3.2, -3.2);

  \draw (8.7,0) -- node[anchor=north] {$\theta_{e}$} (11.8,0);

  \draw[arrow, blue!60, ultra thick] (9.91, -0.75) -- (6.56, 0.57);

  \draw[fill=blue!40] (8, 0) circle (0.5);
  \draw[fill=blue!60] (8, 0) circle (2pt);
  \draw[fill=blue!60] (7.9, 0.1) circle (2pt);
  \draw[fill=blue!60] (8.3, -0.1) circle (2pt);
  \draw[fill=blue!60] (7.7, 0.3) circle (2pt);
  \draw[fill=blue!60] (7.7, -0.15) circle (2pt);
  \draw[fill=blue!60] (7.7, -0.15) circle (2pt);
  \draw[fill=blue!60] (8, -0.3) circle (2pt);
  \draw[fill=blue!60] (8.2, 0.3) circle (2pt);

  \end{tikzpicture}
    \caption{Diagram of a loosely bound dark matter composite entering the Earth, and its subsequent disassembly. The disassembled constituents exit Earth with a spatial spread on the Earth's surface $R_{sp}$.}
    \label{fig:trajectory-diagram}
\end{figure}

In this work, we explore the disassembly of such loosely bound dark matter composites as they cross the Earth, as depicted in Figure \ref{fig:trajectory-diagram}, and outline detection prospects for the disassembled DM constituents in terrestrial DM searches. We characterize the spatial extent, or “spread", of the cascade of disassembled constituents as they exit the Earth with cascade radius $R_{sp}$ on the surface of the Earth, which we find can be a substantial fraction of the Earth's radius.

\begin{figure*}[btp]
    \centering
    \includegraphics[width=\textwidth]{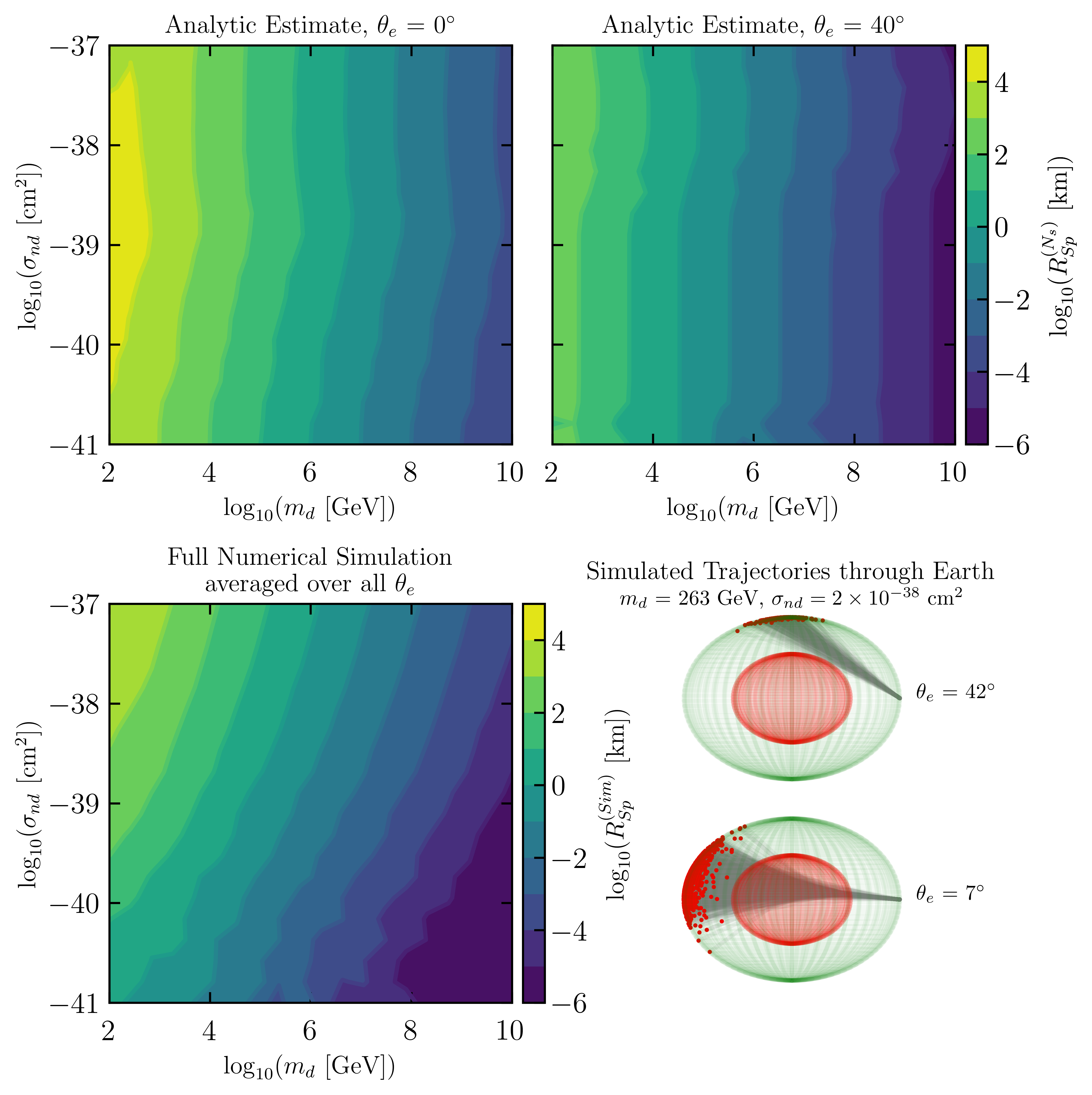}
    \caption{\textit{Top:} Analytical estimates (Eq.~\ref{eq:AnalyticalR}) of the constituent spread radius on the Earth's surface $R_{sp}^{(N_s)}$, in log scale, varying constituent mass $m_d$ and constituent-nucleon cross-section $\sigma_{nd}$, for two entry angles $\theta_e = {0,40}^\circ$. \textit{Bottom left:} Constituent spread radius $R_{Sp}^{(Sim)}$ averaged over flux-weighted entry angle $\theta_e$, using numerical trajectory simulations described in Section \ref{sec:modelling}. \textit{Bottom right}: Example simulated trajectories of $10^3$ dissociated constituents traversing the Earth, for fixed values of constituent mass $m_d$, constituent-nucleon cross-section $\sigma_{nd}$, and entry angles. The final constituent positions upon exiting the Earth are plotted as red points. The Earth's surface$+$mantle and core are respectively shown as green and red spheres.}
    \label{fig:Rspread}
\end{figure*}

We will examine the growth of the disassembled dark matter cascade using analytic formulae and explicit simulation. To begin, let us assume constituents have typical elastic recoils with Standard Model nuclei, and further let us assume that the constituent mass is large relative to the SM nuclear mass, $m_d \gg m_A$. In this case, the constituent's deflection angle in a single scatter with a nucleus is $\sin \theta_d \sim m_A/m_d$. To take into account the Earth's composition, we approximate the average constituent deflection angle from all target nuclei by $\sin \theta_d \sim \bar{m}_A/m_d$, where $\bar{m}_A$ is the mean target mass, which is calculated by weighting each target by its relative abundance and cross-section. Treating successive scatters as a random walk over deflection angles, the total deflection angle after $N_{s}$ scatters is,

\begin{equation}
    \theta_{N_s} \approx \sqrt{N_s} \frac{m_A}{m_d}.
\end{equation}
From this we can obtain the total spread of constituents after $N_s$ scatters (see Appendix \ref{app:Restimate} for some more details),

\begin{equation}
    R_{Sp}^{(N_s)} \approx (\sum_A n_A \sigma_{Ad})^{-1} \sum_{n = 1}^{N_s} \sin\left (\sqrt{n} \frac{\bar{m}_A}{m_d} \right ),
    \label{eq:AnalyticalR}
\end{equation}
where $(\sum_A n_A \sigma_{Ad})^{-1}$ is the mean free path of a constituent in the Earth, summing over all nuclear targets with nucleon number A, assuming a nucleus density of $n_A$ and an interaction cross section $\sigma_{Ad}$.

The rest of this paper is organized as follows. In Section \ref{sec:modelling}, we describe the cosmological formation of loosely bound composite states, and the numerical modelling\footnote{The \texttt{DarkDisassembly} code is available on GitHub at \url{https://github.com/yildab/DarkDisassembly}.} of composite disassembly during transits through Earth, and use this to determine detection prospects for disassembled constituents in Section \ref{sec:detection}. We find that constituent cascades lead to multiscatter detector events in much of the composite parameter space, and describe characteristic time profiles for multiscatter events in detectors in Section \ref{sec:timing}. Since the constituent cascades can be spread over large areas on the Earth's surface, we also consider the possibility of scattering events at multiple detectors in different underground labs in Section \ref{sec:multi-detectors}. In Section \ref{sec:cosmo}, we estimate the fraction of DM particles that would be dissociated from composite states during their cosmological evolution, prior to reaching the Earth. In Section \ref{sec:conclusion}, we conclude. In Appendices, we derive an estimate for constituent cascade spread radii (\ref{app:Restimate}), detail the Earth composition used in simulations (\ref{app:comps}), and show timing profiles for constituent arrivals for different composite entry angles (\ref{app:times}).

\section{Modelling Composite Disintegration in the Earth}
\label{sec:modelling}

We are interested in loosely bound composite dark matter with binding energies per constituent $BE \leq O\text{(keV)}$, so that a typical recoil of a DM constituent with a SM nucleus in a low-redshift galaxy would impart enough energy to eject a constituent from the composite. Such loosely bound composites \cite{Acevedo:2024lyr} arise from, for instance, models of “nuclear dark matter," analogous to SM nuclei, where charged dark fermions form dark nucleons through confinement at a scale $\Lambda_D$, and form dark nuclei through an attractive force sourced by a coupling with a dark pion ($e.g.$ \cite{Hardy:2014mqa,Detmold:2014qqa,Krnjaic:2014xza}). In this case, the binding energy per constituent scales with the inter-constituent spacing $\Lambda_D^{-1}$,

\begin{equation}
    BE(N_D) \approx a_V  \Lambda_D,
\end{equation}
where this expression assumes a simple nuclear liquid drop model with $\mathcal{O}(1)$ volume coefficient $a_V$, as in the SM, although with negligible Coulomb interactions. Another model that could produce loosely bound composites is molecular dark matter, where dark particles form bound states through couplings with a dark electron \cite{Acevedo:2024lyr}. In this case, the interconstituent spacing is set by a dark electron mass and coupling $BE \approx \alpha_{de}  m_{de}.$ The number of constituents expected in a composite $N_D$, can be very large in the case that Coulomb repulsion is absent during nuclear assembly, and also in the case of out-of-equilibrium assembly during, $e.g.$, a first order phase transition \cite{Witten:1984rs,Hardy:2014mqa,Wise:2014ola,Gresham:2017zqi,Bai:2018dqn,Acevedo:2020avd,Acevedo:2021kly,Bleau:2025klr}. In such loosely bound composite scenarios, $\Lambda_D$ and $BE$ can take on a broad range of values, including the energy range $\leq$ keV of particular interest in this work, in which composite disassembly occurs when composites intersect stars and planets. 

In the treatment that follows, we will consider the phenomenological signatures of loosely bound composite dark matter disassembling as it travels through the Earth. Since for these composites, a single scattering event between a dark constituent and a nucleus is sufficient to dislodge the dark particle from the larger composite, it is reasonable to expect that a composite will shed some fraction of particles before reaching the Earth during its cosmological voyage.  In Section \ref{sec:cosmo} we estimate the fraction of free floating particles that result from interactions over the lifetime of the composite, and find this will be less than $10^{-10}$ of the constituents for the parameter space of interest.  

\begin{figure*}[btp]
    \centering
    \includegraphics[width=0.99\textwidth]{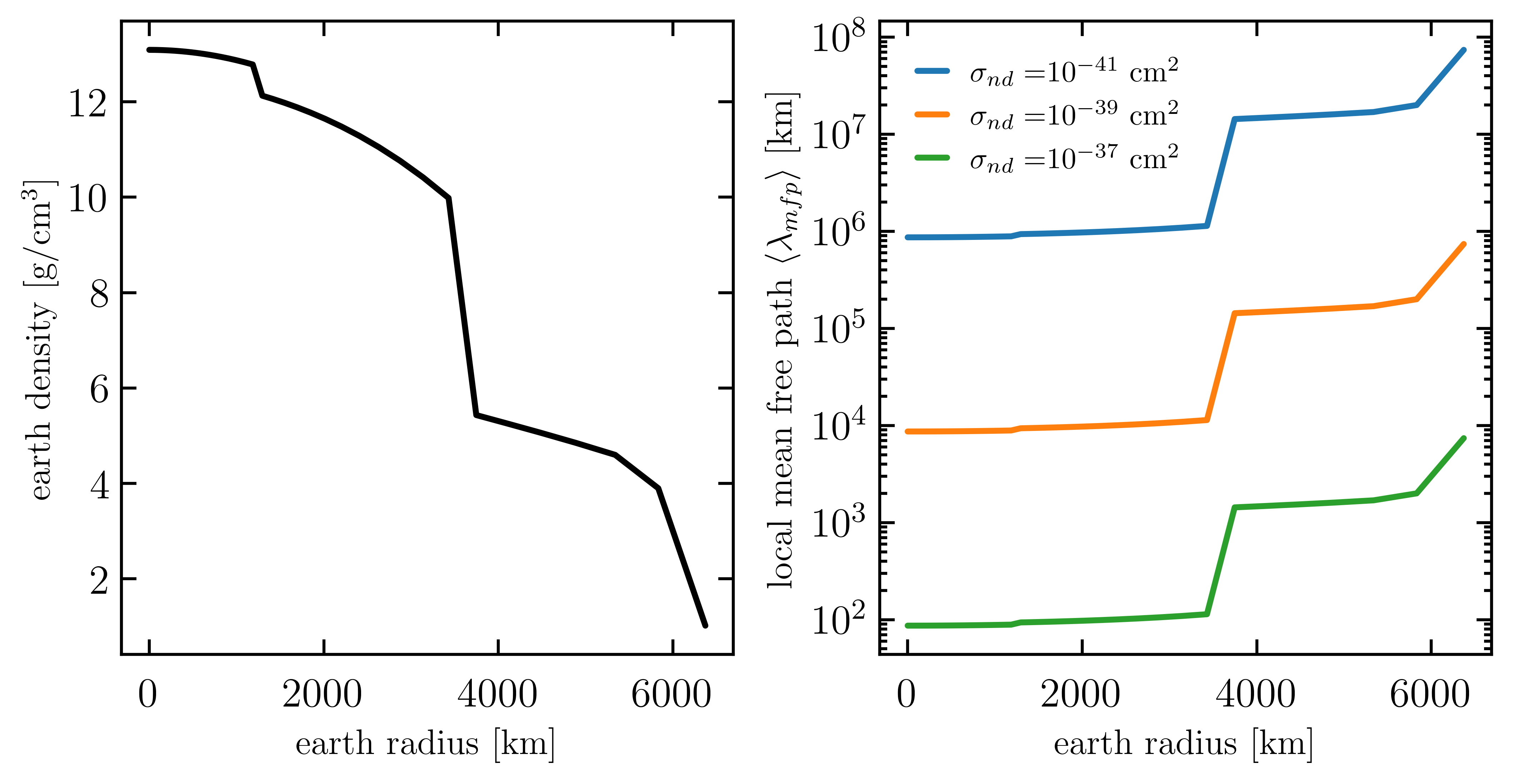}
    \caption{\textit{Left}: The Earth's density as a function of radius from the Earth's centre, according to the Preliminary Reference Earth Model (PREM) \cite{Dziewonski:1981xy}. \textit{Right}: The average local mean free path of a dark matter particle of mass $m_d = 1$ TeV in the Earth, as a function of radius, calculated from the local Earth density and composition, assuming a spin-independent nucleon scattering cross-section $\sigma_{nd}$ for constituent dark matter particles.}
    \label{fig:prem-mfp}
\end{figure*}

\begin{figure*}[t!]
    \centering
    \begin{tikzpicture}[node distance=2cm]
\node (init) [startstop] {Sample entry angle $\theta_{e}$, and initialize $\vec{x}, \vec{v}$.};
% \node (layer) [process, below of=init] {Determine Earth layer.};
\node (mfp) [process, below of=init] {Calculate local mean free path $\langle\lambda_{mfp}(\vec{x})\rangle$, and sample distance travelled until next scatter, $L$.};
\node (checklayer) [question, below of=mfp] {Is the proposed scatter location, $\vec{x} + L\hat{v}$, in the same Earth layer as $\vec{x}$?};
\node (samelayer) [process, below of=checklayer, xshift=-4cm] {Update $\vec{x} \rightarrow \vec{x} + L\hat{v}$. Sample a random scattering target and update $\vec{v}$ from the scattering kinematics.};
\node (difflayer) [process, right of=samelayer, xshift=5cm] {Particle exits layer before scattering. We set $\vec{x} \rightarrow \vec{x} + L_b\hat{v}$, where $L_b$ is the distance to the layer boundary along the particle's trajectory.};
\node (exitcheck) [question, below of=checklayer, yshift=-2cm] {Did the constituent exit the Earth?};
\node (finish) [startstop, below of=exitcheck] {End simulation and save trajectory data.};

\draw [arrow] (init) -- (mfp);
\draw [arrow] (mfp) -- (checklayer);
\draw [arrow] (checklayer) -| node[anchor=east] {Yes} (samelayer);
\draw [arrow] (checklayer) -| node[anchor=west] {No} (difflayer);
\draw [arrow] (samelayer) |- (exitcheck);
\draw [arrow] (difflayer) -| (exitcheck);
\draw [arrow] (exitcheck) -- node[anchor=east] {Yes} (finish);
\draw [arrow] (exitcheck) -- ++(6,0) |- node[anchor=west] {No} (mfp);

\end{tikzpicture}
    \caption{Flowchart showing the numerical method used in \texttt{DarkDisassembly} to model composite dark matter constituent particles disassembling as they traverse the Earth's interior.}
    \label{fig:flowchart}
\end{figure*}
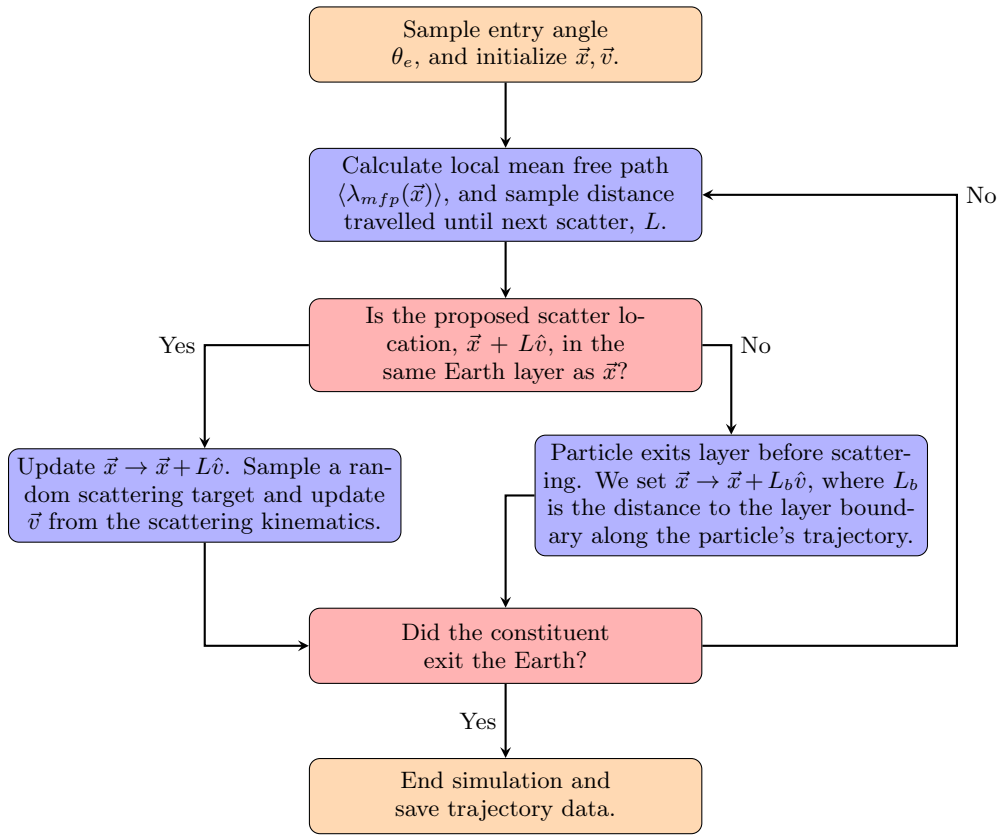

We now turn to the trajectory of loosely bound composites entering and disassembling inside the Earth. We will assume that the dominant scattering regime is constituent-nucleus scattering, as is appropriate for sufficiently large loosely bound composites  \cite{Acevedo:2024lyr}, and that each elastic collision between a constituent and nucleus imparts enough energy, so that constituents are ejected from the composite after scattering. For much of the parameter space explored in this work, we simulate trajectories for a sample population of $N_{sim}  = 1000$ individual dark matter constituent particles passing through the Earth, for a constituent mass and nucleon cross-section denoted by $m_d,~\sigma_{nd}$.

For both simulations and analytic estimates, the Earth is modelled as a sphere composed of three concentric regions (the core, the mantle, and the crust) with distinct element compositions, following the treatment in \cite{acevedo2023oldrocksnewlimits, Acevedo_2021,Bramante_2020}. The density of the Earth, as a function of radius, is modelled according to the Preliminary Reference Earth Model (PREM) \cite{Dziewonski:1981xy}, shown in the left panel of Figure \ref{fig:prem-mfp}. The elemental compositions used are given in Table \ref{tab:elements} in Appendix \ref{app:comps}.

For the results presented in this study, we model each of the $N_{sim}$ constituent trajectories starting at the Earth's surface, and set their speed to $v = 10^{-3} c$. We sample the composite's entry angle $\theta_{e}$ with respect to the normal to the Earth's surface using a probability distribution,
\begin{equation}
    \frac{P(\theta_e)}{d\theta} = 2 \sin\theta_e\cos\theta_e, ~ \theta_e \in \left [0, \frac{\pi}{2}\right],
    \label{eq:probangle}
\end{equation}
which is appropriate for an isotropic flux onto a spherical surface. The entry angle $\theta$ determines the initial direction of a composite (and its constituents) with velocity vector $\hat{v}$. The local mean free path of the constituent is calculated by summing over all nuclear targets with nucleon number $A$~\cite{emken2019darkmatterearthsun},

\begin{equation}
    \langle \lambda_{mfp} (\vec{x}) \rangle = \left( \sum_A n_{A}(\vec{x}) \sigma_{Ad}\right)^{-1}.
\end{equation}

The local mean free path as a function of radius for some characteristic values of $\sigma_{nd}$ is plotted in the right panel of Figure \ref{fig:prem-mfp}. From the mean free path, the code samples the distance travelled by the constituent before it scatters through inverse transform sampling,

\begin{equation}
    L = -\ln(1-\zeta)\lambda_{mfp}, ~\zeta \in \mathcal{U}(0,1),
    \label{eq:Lscatter}
\end{equation}
where we take $\zeta$ to be equally weighted across the interval $(0,1)$, and so this equation assumes that the mean free path $\lambda_{mfp}(\vec{x})$ does not vary over the distance $L$. Modelled this way, the constituent will therefore scatter at $\vec{x_i} = \vec{x} + L\hat{v}$. From the relative elemental abundances at $\vec{x_i}$, the code randomly selects a scattering target nucleus, weighting the random selection of target nucleus according to relative abundance and cross-section $\sigma_{Ad}$, and determines the constituent's energy loss and elastic scattering angle. Using this the code updates $\vec{v}$. The mean free path is calculated at the new location, and the process is repeated until the constituent re-exits the Earth. Figure \ref{fig:flowchart} depicts a flowchart of this numerical process.

As noted above, Equation \ref{eq:Lscatter} assumes the mean free path is fixed as function of radius over scattering length scales. This is an appropriate simplification within a particular Earth layer, as we see in Equation \ref{fig:prem-mfp}, but something different happens if the constituent crossed into a new layer before scattering at $\vec{x_i} = \vec{x} + L\hat{v}$. If the prescription in Eq.~\eqref{eq:Lscatter} results in the constituent crossing into a new layer ($i.e.$ crust, mantle, core) before scattering, the code updates the constituent trajectory to $\vec{x_i} = \vec{x} + L_b\hat{v}$, where $L_b$ is the distance to the new layer. Then the next scattering length is computed using the local mean free path according to the new layer's composition and density. This procedure yields final locations for constituents on the Earth's surface after their transit, from which we determine the radius of the constituent cascade spread on the surface, $R_{Sp}$. For numerical results we formally define $R_{Sp}$ as the distance from the centre of the spread within which 90\% of the constituents lie, for a given $\theta_{e}$. To obtain results shown in this study, we have averaged over twenty $\theta_{e}$ orientations, weighted according to \autoref{eq:probangle}, in order to get a flux-weighted spread radius. 

In the bottom left plot of Figure \ref{fig:Rspread}, we show the spread radius as a function of constituent-nucleus cross-section and constituent mass. The size of the constituent spread across the Earth's surface is largest at higher cross-sections, due to higher scattering rates, as one might expect. The spread size also decreases with increasing constituent mass, as the deflection angle per scatter is inversely proportional to $m_d$, $\theta_d \simeq m_A/m_d$. We also highlight that $R_{Sp}$ is independent of the number of constituents in the composite, as long as we remain in the loosely bound regime.

\section{Direct Detection Prospects for Disintegrated Composites}
\label{sec:detection}

The phenomenological signatures of a “cascade" of constituents entering a detector can be rather different than a constant flux of individual DM particles. We find that for a swathe of relevant constituent parameters, $N_D, \sigma_{nd}, m_d$, the passage of a constituent cascade can lead to multiple scattering events in a detector, requiring dedicated multiscatter dark matter searches. 

The number of scatters expected in a detector for a constituent cascade of radius $R_{Sp}$ is
\begin{equation}
    N_{sc/Sp} = n_A \sigma_{Ad} l_{det} \ N_D \min\left (1, \frac{A_{det}}{\pi R^2_{Sp}}\right), 
    \label{eq:Nscatterscone}
\end{equation}
where $l_{det},~A_{det}, n_A$ refer to the detector's length, area, nuclear number density, and the final factor takes into account the fraction of constituents we expect to traverse the detector. The constituent-nucleus cross-section $\sigma_{Ad}$ is calculated from the differential cross-section,
\begin{equation}
    \frac{d\sigma_{Ad}}{dE_R} = \left(\frac{\mu_{Ad}}{\mu_{nd}}\right)^2A^2\frac{d\sigma_{nd}}{dE_R}|F_A(q)|^2\,
\end{equation}
where $F_A(q)$ is the nuclear Helm form factor \cite{Helm:1956zz}. The maximum value of the recoil energy $E_R$ for nuclear scattering in the detector is set by the constituent's velocity upon arriving in the detector. 

The number of times that a constituent spread is expected to pass through a $\sim m^2$-sized detector in a year is equal to the flux of composites through the Earth in a year, times the probability that a given composite's constituent spread will go through a detector, which we can relate to the fraction of the Earth's surface taken up by the spread (or by the detector itself, if the size of the spread is smaller than $A_{det}$). Then the rate for cascade spreads hitting a detector is
\begin{equation}
    \dot{N}_{Sp/det} \simeq \dot{N}_{comp/\oplus}\times \max\left(\frac{\pi R_{Sp}^2}{4\pi R_{\oplus}^2}, \frac{A_{det}}{4\pi R_{\oplus}^2}\right),
    \label{eq:ncloudsyear}
\end{equation}
where $\dot{N}_{comp/\oplus} =\frac{1}{4} n_D v_d 4 \pi R_{\oplus}^2 $ is the flux of dark composites through the Earth, with composite number density in the Milky Way halo $n_D = \rho_d/(N_D m_d)$ for $\rho_d \sim 0.3~{\rm GeV/cm^3}$ and $v_d \sim 10^{-3}$ the average dark matter speed in the halo. Combining Equations \ref{eq:Nscatterscone} and \ref{eq:ncloudsyear}, the rate of individual constituent-nucleus scatters in a detector of area $A_{det}$ is
\begin{equation}
    \dot{N}_{sc/det} = \frac{1}{4} n_A \sigma_{Ad} l_{det}  n_{d} v_d A_{det},
    \label{eq:Nscatterpyear}
\end{equation}
where $n_d = \rho_d /m_d$ is the number density of dark matter constituents in the halo.

\begin{figure}
    \centering
    \includegraphics[width=\linewidth]{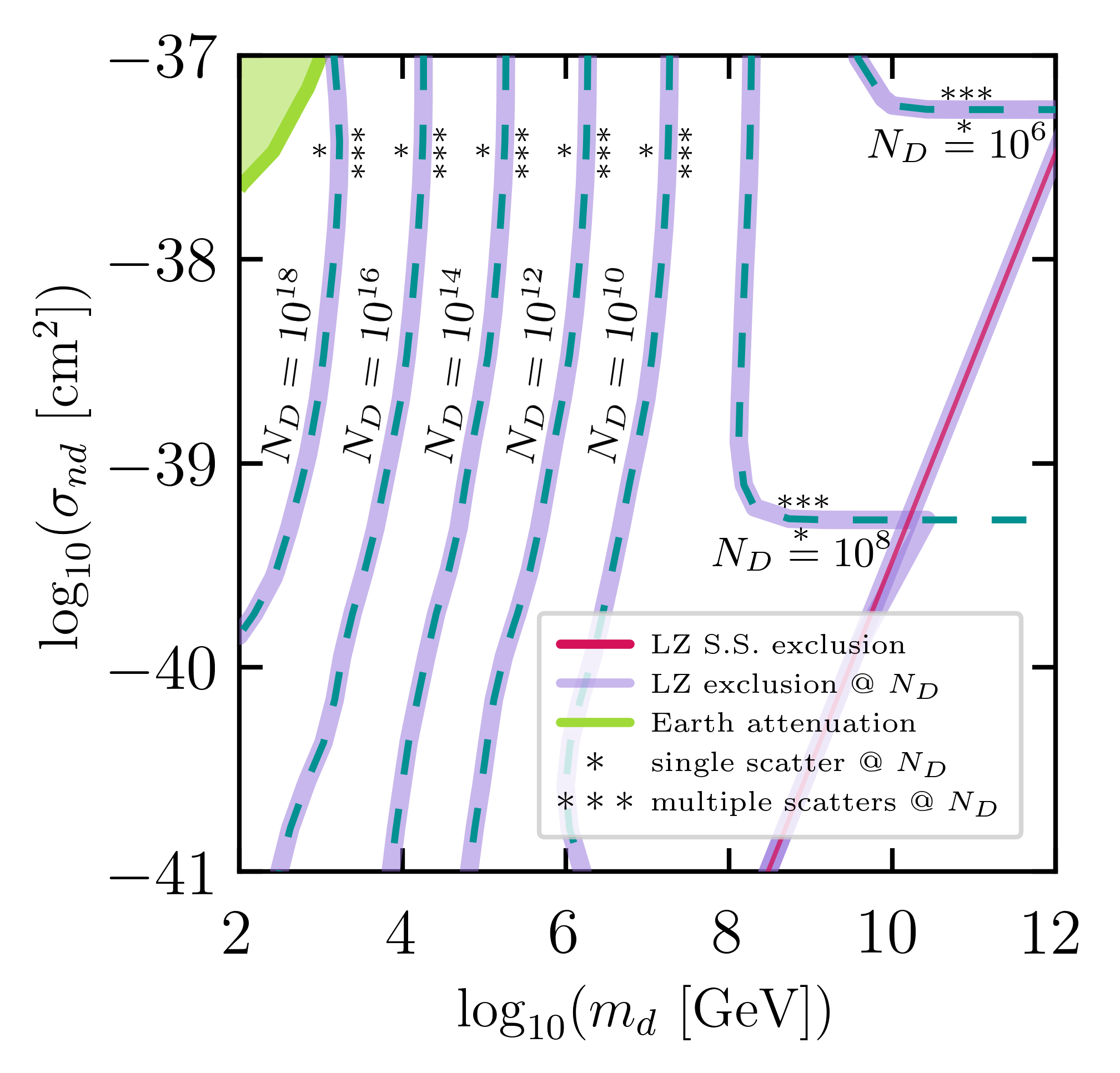}
     \caption{Direct detection constraints and prospects for disassembling composite dark matter, for constituent mass $m_d$ and constituent-nucleon cross-section $\sigma_{nd}$. The regions to the left of the purple lines marked for $N_D=10^{18}-10^{6}$ constituents are estimated exclusions for single scatter searches at experiments like LUX-ZEPLIN (LZ), XENON-NT and PANDA-X, and DEAP-3600, which would have seen single scatter dark matter events given the spread of constituents. We calculate these estimated exclusions by assuming a target volume of a cubic metre of liquid xenon of density $3.52$ g/cm$^3$ with a year observation time. The regions to the right of these purple lines require new multiscatter searches discussed in the text, because a typical cascade produces multiple coincident recoils in underground detectors. For comparison, the solid red line shows the published constraint on spin-independent dark matter from LZ, extrapolated to higher dark matter masses \cite{Aalbers_2025}. The teal dashed contours, which are often coincident with the purple single scatter bounds, indicate the separation between parameter space for single-scattering vs multi-scattering due to DM cascades, for constituent number $N_D$ indicated. The green shaded region delineates the Earth attenuation region, where at least 50\% of the individual constituents have lost more than 90\% of their energy due to scattering in the Earth before reaching the detector.}
    \label{fig:constraints}
\end{figure}

Figure \ref{fig:constraints} shows single scatter direct detection constraints and prospective multiscatter search parameter space for DM composites disassembling in the Earth before reaching a detector. We find that a typical cascade of DM constituents from a single disassembled composite is often expected to cause multiple scatters in a detector, particularly in the high cross section region, and for heavier constituent masses, for which $R_{Sp}$ is smaller, implying a higher density of constituents passing through the detector. Higher values of $N_D$ also sizably increase the constituent density through a detector, resulting in multiple scatters even for TeV+ mass constituents.  Notably, we see that ``WIMP-like" TeV-scale dark matter with weak-scale cross sections, for which underground experimental searches are usually particularly sensitive, would not be excluded by direct detection experiments if such WIMP-like constituents exist within very loosely bound composites of $N_D \sim 10^{14}-10^{18}$ constituents.

\begin{figure*}[btp]
    \centering
    \includegraphics[width=\textwidth]{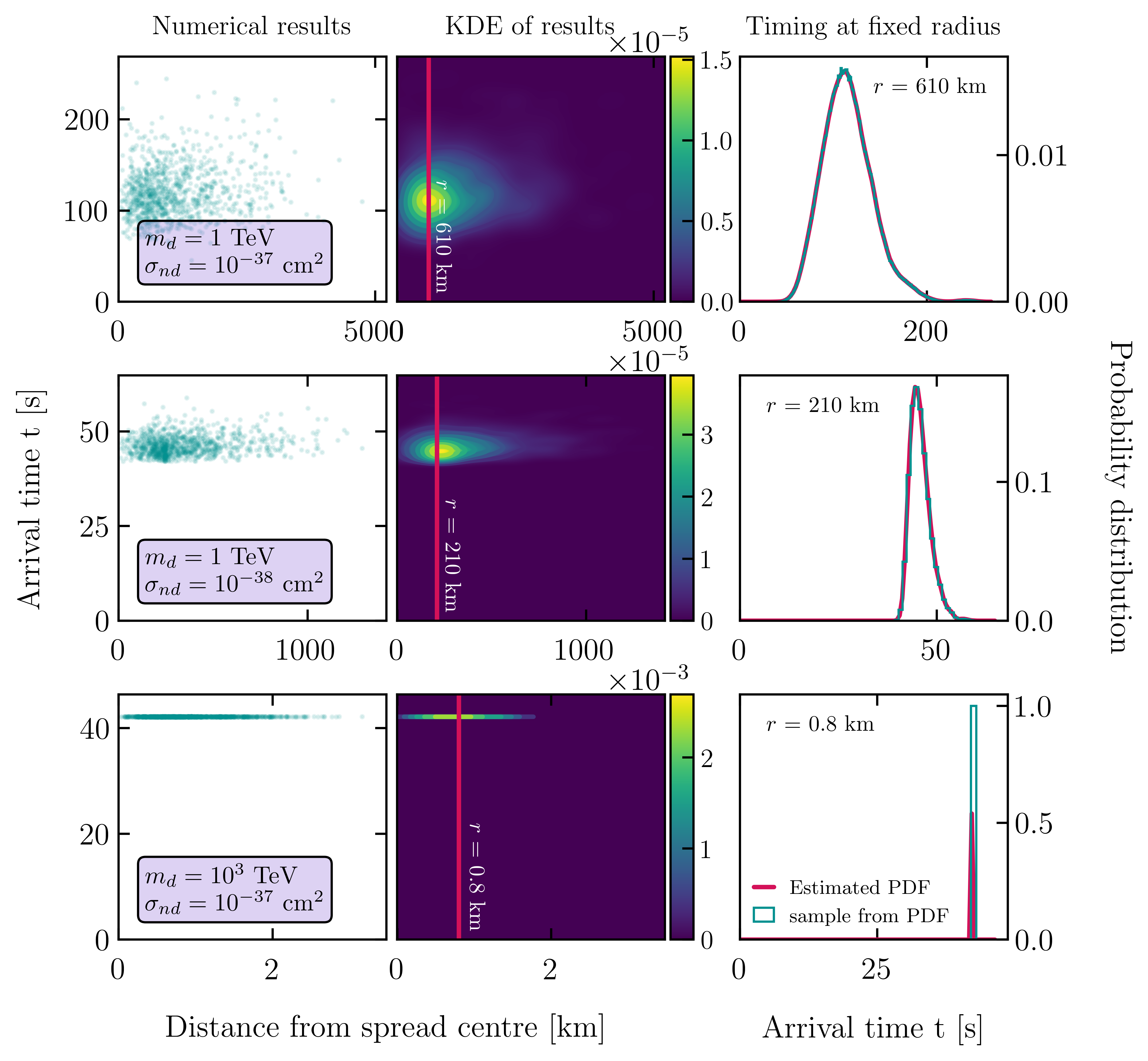}
    \caption{\textit{Left}: constituent arrival time for varying radius from the center of the cascade spread, obtained from a numerical simulation of $N_{sim = }10^3$ constituents and entry angle $\theta_e = 7$ degrees, for different constituent mass $m_d$ and constituent-nucleon cross section $\sigma_{nd}$ indicated in each row. \textit{Middle}: Kernel density estimator of the same dataset, showing the relative probability of constituent arrival times for different radii. \textit{Right}: Probability distribution of composite arrival times for a fixed radius $r$ from the spread's centre, corresponding to the radius at which the constituent density is highest, also shown as the vertical line in the central figure.}
    \label{fig:timingstats}
\end{figure*}

\section{Characterizing constituent arrival times}
\label{sec:timing}

\begin{figure*}[btp]
    \centering
    \includegraphics[width=\textwidth]{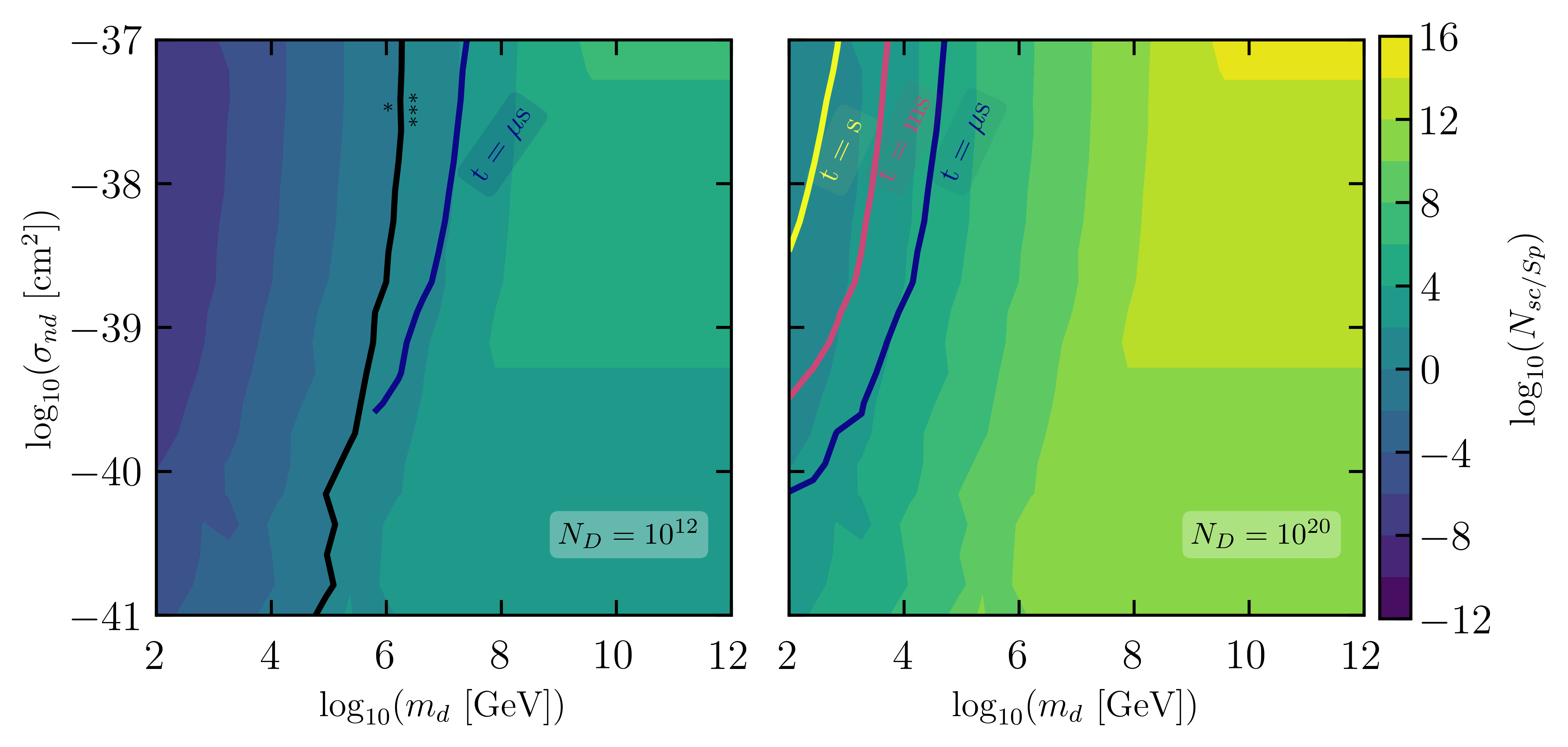}
    \caption{Number of expected constituent scatters per composite cascade passing through 1 m$^3$ of liquid Xenon, for $N_d = 10^{12}$ constituents (left) and $N_D = 10^{20}$ constituents (right). The black contour separates the single-scatter region from the multiscatter region, and the coloured contours indicate the expected time delay between successive scatters in a multiscattering event.}
    \label{fig:scatterpcone}
\end{figure*}

The results presented above motivate multiscatter event searches as part of dark matter experimental search regimes. Multiscattering searches have already been undertaken by the DEAP-3600, XENONnT and LUX-ZEPLIN (LZ) collaborations \cite{DEAPCollaboration:2021raj, XENON:2023iku,LZ:2024psa}, in the case of ultraheavy multiply-interacting massive particles (MIMPs) \cite{Bramante:2018qbc,Bramante:2018tos,Bramante:2019yss,Acevedo:2020avd,Acevedo:2021kly,Acevedo:2024lyr}. For very heavy single-state dark matter considered in the MIMP regime, successive scattering events are expected to be collinear along the individual particle's trajectory. On the other hand, multiscattering due to disassembled composites can manifest as many DM particles scattering in a detector in close succession, without any particular geometry. This requires a different analysis and characterization of the expected spatial and temporal distribution of DM events across the composite parameter space. The discussion in Section \ref{sec:modelling} of the constituent spread is a starting point for determining the expected spatial distribution of scatters in a detector. 

In this Section, we begin exploring the expected time delay between successive constituent-nucleus scattering events in dissociated constituent cascades. To explore this, we extract the time taken for constituents to re-exit the Earth for each point in the ($m_d,~\sigma_{nd}$) parameter space from the trajectory and velocity data generated in the simulated data. In the leftmost panels of Figure \ref{fig:timingstats}, we show the constituent arrival time versus distance from the centre of the constituent spread, with each row of panels sampling a different mass and cross-section ($m_d,~\sigma_{nd}$), as indicated. For that plot, we fix the composite entry angle to $\theta_e = 7$ degrees, so that the constituents roughly traverse the entire Earth before re-exit (see Appendix \ref{app:times} for a wider range of entry angles). As one might expect, heavier constituents ($m_d = 10^3$ TeV, in the third row) have a very narrow spread in arrival time: the energy lost through scattering does not appreciably slow them down, and they are less deflected by these scatters. Hence in this case, all constituents travel roughly the same distance to reach the detector. We thus see very little deviation in arrival times from the Earth crossing time of a dark matter particle passing directly through the centre of the Earth, absent any scatters, which is roughly 42 seconds. On the other side hand, TeV-scale constituent particles are substantially slowed by successive scatterings in the Earth, with arrival times spread over up to hundreds of seconds.

We estimate the probability density function of this two-dimensional space using numerical kernel density estimation (KDE) from the Python package \texttt{kdetools}, and plot this probability density function in the second column of Fig. \ref{fig:timingstats}. Fixing the distance $r$ to the centre of the constituent spread, we recover the probability density function of constituent arrival times, as seen in the third column of panels in Figure \ref{fig:timingstats}.

Using this framework, we sample the arrival times of constituents expected to scatter within a direct detection experiment, calculated from Eq. \ref{eq:Nscatterscone}, and use this to determine the time delay between successive scatters in the target volume due to a single constituent cascade. In Fig. \ref{fig:scatterpcone}, we show the number of expected scatters per cascade of constituents passing through a cubic metre of liquid xenon, for two example choices of $N_D$. In the multiscattering region, the coloured contours show the expected timing between successive scattering events, which ranges from microseconds to seconds in WIMP-like dark matter parameter space, for $N_{D} =10^{12},10^{20}$ constituents. Comparing the two panels of Figure \ref{fig:constraints} shows that a larger number of constituents $N_D$ leads to a larger number of expected scatters within the target volume, and thus a smaller expected time delay between scattering events on average. 

We have particularly shown contours for inter-event times ranging from seconds to microseconds in order to highlight parameter space where direct detection experiments would plausibly have the timing resolution necessary to resolve individual scattering events. The time delay between events, combined with position reconstruction, can be a useful additional handle to distinguish this signal from possible Standard Model backgrounds, such as neutron events. For time delays smaller than microseconds, we expect consecutive scatters will pile up and appear as simultaneous energy deposition across different positions in the target volume, or saturate the detector at some upper threshold. The expected multiscattering signal thus depends greatly on the particulars of the detector response to a large number of scatters.

\section{Correlated events at different detectors}
\label{sec:multi-detectors}

\begin{figure*}[btp]
    \centering
    \includegraphics[width=\textwidth]{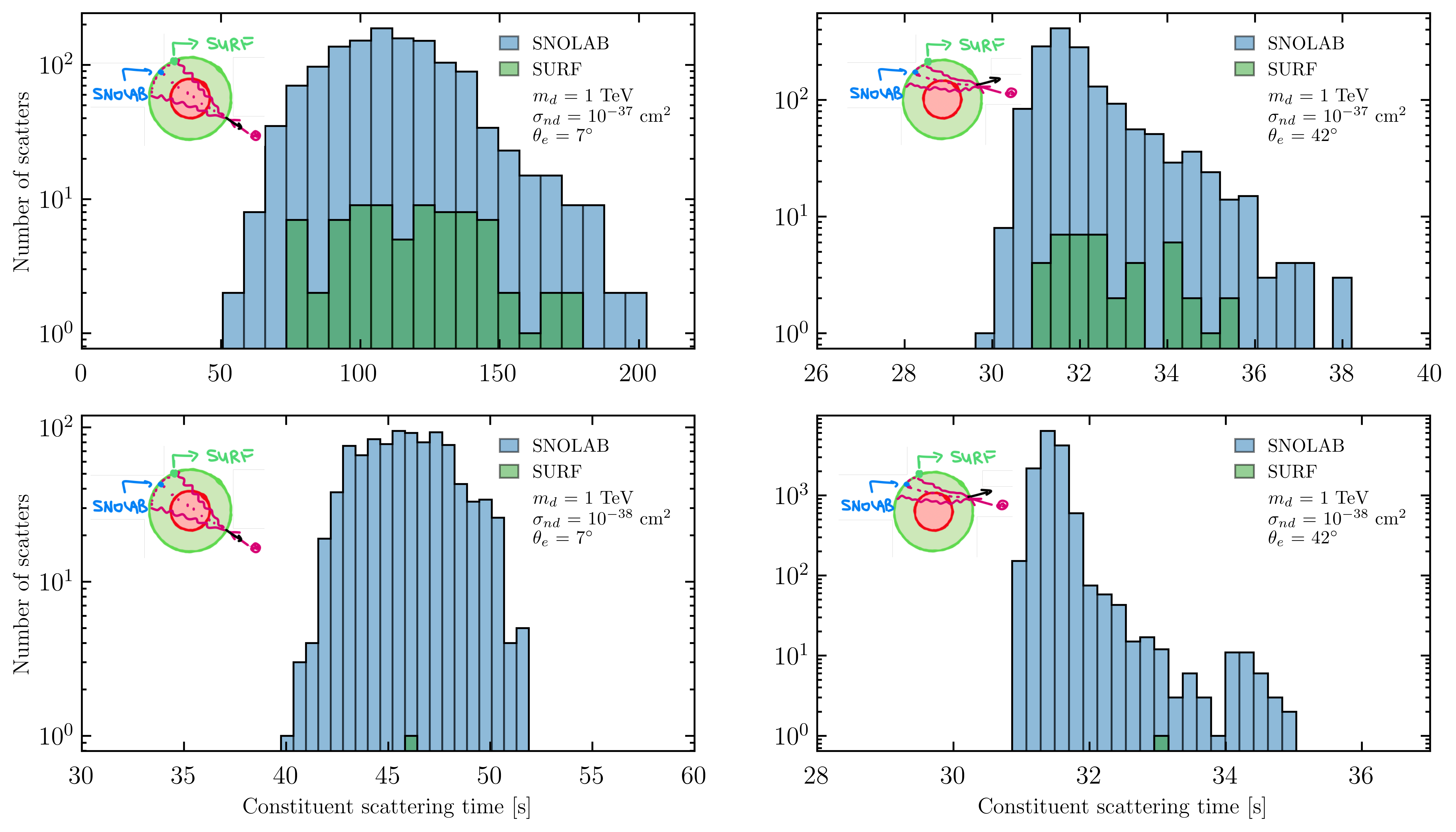}
    \caption{Arrival times of constituents from the same constituent cascade traversing both SNOLAB and SURF laboratories, varying the constituent-nucleis cross section and the composite entry angle $\theta_{e}$. We assume here that SNOLAB is located at the centre of the cascade and SURF is located 1700 km radially away from this point, as pictorially shown.}
    \label{fig:separatedetectors}
\end{figure*}

In earlier sections, we found that cascades of dissociated constituents can have a large spatial extent on the Earth's surface, up to thousands of kilometres. This motivates exploring an interesting signal wherein a single cascade of constituents traverses two detectors, and leads to correlated events across them. In this Section, we sketch out what such a signal may look like.

We consider constituents with mass $m_d = 1$ TeV and cross-sections with nucleons of $\sigma_{nd} = 10^{-37}$ cm$^2$ and $\sigma_{nd} = 10^{-38}$ cm$^2$. This is a region in parameter space where the size of the constituent spread is as large as possible, which corresponds to high constituent-nucleon cross-sections and low constituent masses, while still remaining just outside of the Earth attenuation region in Figure \ref{fig:constraints}. At this particular point, the constituent spread radius is approximately $10^3$ km. We therefore select two research laboratories separated by a similar distance: the Sudbury Neutrino Observatory (SNOLAB), in Sudbury ON, and the Sanford Underground Research Facility (SURF) in Lead, South Dakota, which are approximately 1700 km apart. We arbitrarily consider the case where SNOLAB is located at the centre of the constituent cascade, and SURF closer to the edge of the cascade. 

Using the same methodology described in Section \ref{sec:timing}, we determine the number of scatters expected in a detector consisting of one cubic metre of liquid xenon, located at each of these two locations within the constituent cascade, and their expected scattering times (where $t = 0$ corresponds to the time at which the original composite first entered the Earth). We compare results from composites with an initial entry angle that is close to head-on ($\theta_{e} = 7^\circ$) and a shallower entry angle ($\theta_{e} = 42^\circ$). We show the expected distribution of events in Figure \ref{fig:separatedetectors}.

We see that scattering events occur over a larger time window for constituents with larger interaction cross-sections, which are more slowed and deflected by their traversing the Earth. The spread of times also depends greatly on the original composite's entry angle, as a steeper angle leads to constituents traversing a larger distance within the Earth and undergoing more scatters and deviations. We also find fewer scatters, and a smaller scattering time window, for SURF, the outlying detector, compared to SNOLAB, the detector positioned closer to the centre of the spray, as the constituent flux drops off as one approaches the exterior regions of the cascade. Only at large cross-section can we expect to see multiple scatters occur within the outlying detector.

\section{Cosmological dissociation of loosely bound composites}
\label{sec:cosmo}

We would like to know what fraction of constituents we should expect to dislodge from loosely bound composites, after composite formation, prior to hitting the Earth. This will determine whether it would be better to simply search for this sub-population of dissociated constituents, rather than search for the large $N_D$ composites that more rarely cross the path of the Earth. 

Below we consider a number of processes that would dissociate composites, including scattering with nuclei prior to galaxy formation, scattering with gas in the Milky Way, hitting stars in the Milky Way, hitting cosmic rays, and being disrupted by tidal forces either from passing near a star or from forming a binary system of two composites that inspiral toward each other. We find that collisions with cosmic rays and tidal interactions break up a negligibly small fraction of composites as compared to the other disruption mechanisms.  We will focus the discussion below on disruptions caused by collisions with gas in the Milky Way, including the epoch prior to the Milky Way's formation, and collisions with stars as these have the largest effect by many orders of magnitude.

Collisions between composites and stars in the Milky Way disrupts the largest fraction of composites with cross sections below $10^{35}$ cm$^{2}$.  The column density of nuclei in a star is as large as or greater than that of the Earth.  Any composite we expect to disassemble when passing through the Earth would also fully disassemble if it hit a star.

To estimate what fraction of composites have hit a star, we assume all stars have an average size and estimate the probability of hitting a star of that size within 10 Gyr.  To find typical properties of stars in the Milky Way such as average mass, total number of stars and the average target size of a star we assume the they are distributed according to a Salpeter Initial Mass Function.   \cite{1955ApJ...121..161S}.
\begin{equation}
\label{salpeter}
    \frac{dN}{dM} \propto M^{-2.35}
\end{equation}
We integrate this over a range of possible stellar masses, $0.085 M_{\odot}-120M_{\odot}$ based on the masses of the heaviest \cite{1997ARA&A..35....1D} and lightest \cite{von_Boetticher_2017} stars observed in the galaxy, to find the average stellar mass in the Milky Way. We found this to be about 0.3 $M_{\odot}$. Taking the total mass of stars in the Milky Way to be about $6\times 10^{10} M_{\odot}$ \cite{McMillan_2011}, we estimate that there are $\sim 2\times10^{11}$ stars in the galaxy. 
Using Eq.\ref{salpeter} and the stellar mass-radius relation \cite{2021foas.book.....R}
\begin{equation}
    \begin{cases}
        R/R_{\odot}=1.06 (M/M_{\odot})^{0.945} &M<1.66M_{\odot}\\
         R/R_{\odot}=1.33 (M/M_{\odot})^{0.555} &M>1.66M_{\odot}
    \end{cases}
\end{equation}
we find the average cross sectional area of a star in the Milky Way to be $\sim0.3\pi R_{\odot}^2$.  We use this as the average star-composite cross section.  The probability of scattering with a star in 10 Gyrs is then roughly
\begin{equation}
    f_{scatter} = \frac{N_{*}\sigma_{*}v_{comp}t}{V_{halo}}\sim4\times 10^{-11}
\end{equation}
where $t=10$ Gyrs, $N_{*}=2\times10^{11}$ is the number of stars in the Milky Way, $\sigma_{*} = \pi R_{\odot}^2$ is the average cross sectional area of a star in the Milky Way, $v_{comp}= 10^{-3}$ is the typical velocity of the composite, and $V_{halo}=4\pi(4\times 10^5\text{lyrs})^3/3 $ is the volume of the dark matter halo of the Milky Way. Assuming composites are fully disassembled after hitting a star, we expect a fraction of about $\sim 4\times 10^{-11}$ of all dark constituents to be free floating due to collisions with stars (put another way, we expect fewer than $4 \times 10^{-11}$ composites to have hit a star).

Collisions between composites and particles of the interstellar medium (ISM) also contribute to the population of free floating constituent particles. Composites move through the ISM  at a velocity of $10^{-3}$, fast enough that any collision between a composite and an atom should be energetic enough to dislodge a dark particle. Assuming the ISM is composed primarily of Hydrogen nuclei, and has a mass equivalent to 10$\%$ of the stellar mass of the Milky Way \cite{Scoville_2016}, then one would expect a fraction $f=M_{ISM}\sigma_{nd}v_{comp}t/V_{halo}\sim2\times 10^{-11}\frac{\sigma_{nd}}{10^{-35}\text{cm}^2}$ of all particles in a composite to hit a Hydrogen atom over the $\sim$10 billion year lifetime of the Milky Way. Here $M_{ISM}$ is the mass of the material in the ISM.

Composites may also scatter with nuclei and lose component particles prior to galaxy formation.  In the time between composite formation and galaxy formation the Universe would be more hot and dense than it is today, increasing the scattering rate.  We estimate the fraction of dislodged particles by integrating the probability of a particle in a composite scattering with an H atom from between when the composite would have formed, $z\sim BE/T_0$, up to the time of galaxy formation.  Here we assume the Milky Way formed at z=6, but this estimate does not change if we assume galaxy formation happens at a much earlier time such as z=20. Most of the scattering would take place at the highest redshifts where greater temperatures and densities increase the probability of a particle scattering energetically enough with H to separate from the composite. 
\begin{equation}
    \begin{split}
        f_{free}=&\sigma_{nd}\rho_{crit}\Omega_b\\\times&\int_{BE/T_0}^6\frac{(1+z)^2f_{v}(z,BE) dz}{H_0\sqrt{\Omega_m(1+z)^3+\Omega_{\gamma}(1+z)^4+\Omega_{\Lambda}}}\\
        \sim& 2\times 10^{-15}\left(\frac{\sigma_{nd}}{10^{-35}\text{cm}^2}\right)\left(\frac{BE}{10^{-6}\text{GeV}}\right)^{2.36}
    \end{split}
\end{equation}
where $H_0$ is the Hubble constant today, $\Omega_m$, $\Omega_{\gamma}$, $\Omega_{\Lambda}$, and $\Omega_b$  are the matter, radiation, dark energy and Baryon fractions of the Universe respectively, $T_0$ is the radiation temperature today, $\rho_{crit}$ is the critical density of the Universe. The $f_v(z,BE)$ term is the fraction of nuclei with a momentum large enough to dislodge a particle, $p>BE$ at a given redshift $z$ times the velocity of said particle.  It can be found by taking
\begin{equation}
    f_v(z,BE)=\frac{1}{n}\int_{BE}^{\infty}dp^3\left[e^{(1+p^2)^{1/2}/T_0(1+z)}+1\right]^{-1}v(p)
\end{equation}
where for a Hydrogen nucleus, taking $m_p=1$, the Hydrogen velocity is $v(p)=\frac{p}{\sqrt{1+p^2}}$.  The term $n=\int_{0}^{\infty}dp^3\left[e^{(1+p^2)^{1/2}/T_0(1+z)}+1\right]^{-1}v(p)$ is a normalizing factor that comes from integrating over the entire range of $p$. While this contribution to the fraction of free particles is small compared to scattering \textit{within} the Milky Way, it can become dominant for large values of $BE$.  For the models considered in this work, the free floating fraction of particles is never large enough to be itself constrained by current dark matter searches.

\section{Conclusion}
\label{sec:conclusion}

In this work, we have investigated distinct terrestrial signatures of loosely bound composite dark matter. Unlike point-like or tightly-bound composite dark matter, loosely bound composites disassemble upon interaction with the Earth's interior, creating a wide cascade of constituents. With both analytical estimates and detailed numerical simulations, we have seen that this disassembly leads to a significant spatial cascade spread $R_{sp}$, extending up to thousands of kilometers, with characteristic cascade timing profiles that depend on composite parameters.

We have found several implications for current and future dark matter searches. First, much of the composite parameter space predicts multiple non-collinear scattering events in a single detector. This suggests that loosely bound composites may have evaded both standard single-scatter exclusion limits, and multiscatter limits that assumed a collinear set of scatters \cite{DEAPCollaboration:2021raj,XENON:2023iku,LZ:2024psa}, necessitating a different kind of dedicated multiscatter search analysis. Second, the time delay between successive constituent scatters, which can range from microseconds to seconds, provides a powerful handle for background discrimination and parameter estimation. Furthermore, for large constituent spreads, a single composite entry can trigger correlated signals in geographically separated underground laboratories, such as SNOLAB and SURF, offering a unique global detection signature. Finally, we estimated that these composites remain largely intact during their cosmological evolution, with less than $10^{-10}$ of constituents dissociating before reaching Earth, meaning that terrestrial disassembly signatures are the primary discovery channel.

Future work should integrate constituent cascade models directly into experimental detector simulations to precisely quantify efficiencies for multiscatter reconstruction. In addition, the analysis of composite dissassembly presented here could be extended to intermediate binding energy regimes where composites may undergo inelastic excitations or partial fragmentation rather than total disassembly, potentially yielding distinct detection signatures. Exploring the detection prospects for these scenarios across a wider variety of composite models remains a compelling avenue for further investigation.

\section*{Acknowledgements}
We acknowledge support from the Natural Sciences and Engineering Research Council of Canada (NSERC), the Ontario Early Researcher Award (ERA), and the Canada Foundation for Innovation (CFI). This research was undertaken thanks in part to funding from the Arthur B. McDonald Canadian Astroparticle Physics Research Institute. Research at Perimeter Institute is supported by the Government of Canada through the Department of Innovation, Science, and Economic Development, and by the Province of Ontario.

\bibliographystyle{JHEP}
\bibliography{biblio}

\newpage
\newpage

\appendix

\onecolumngrid
\newpage
\appendix

\section{Analytical estimate of the spread size after $N_s$ scatters}
\label{app:Restimate}
We can treat the problem of estimating the spread size $R_{sp}$ for constituents scattering in the Earth, as a random walk in two dimensions, such that $\hat{x}$ is the initial heading of the DM particle prior to scattering, and $\hat{y}$ is perpendicular to this. Then we can obtain the expected displacement in the $\hat{y}$ direction after $N_s$ scatters, corresponding to the radius of the conical spread at this point. To do this, we sum successive scattering angles in order to determine the total angle. This is a one dimensional random walk, where the walk steps are equal to $\pm\theta_{scatter}$ so that after $N_s$ scatters the expected deflection angle is $\braket{\theta_n} = \sqrt{N_s} \theta_{scatter} \approx \sqrt{N_s} \frac{m_A}{m_d}$. Then the total displacement in $\hat{y}$ direction is simply the sum

\begin{equation}
    R_{sp} \simeq (n_A \sigma_{Ad})^{-1}\sum_{n = 1}^{N_S}   \sin\left(\sqrt{n} \frac{m_A}{m_d}\right)
\end{equation}

\section{Elemental composition of Earth}
\label{app:comps}

\renewcommand{\arraystretch}{1.2}% for the vertical padding
\begin{table*}[h!]
\centering
\begin{tabularx}{\textwidth}{l|YYYYYYYYYYYYY}
\hline
 & $^{16}$O & $^{28}$Si & $^{27}$Al & $^{56}$Fe & $^{40}$Ca & $^{23}$Na & $^{39}$K & $^{24}$Mg & $^{48}$Ti & $^{57}$Ni & $^{59}$Co & $^{31}$P & $^{32}$S \\ \hline
Crust \%w/w & 46.7 & 27.7 & 8.1 & 5.1 & 3.7 & 2.8 & 2.6 & 2.1 & 0.6 & - & - & - & - \\ \hline
Mantle \%w/w & 44.3 & 21.3 & 2.3 & 6.3 & 2.5 & - & - & 22.3 & - & 0.2 & - & - & - \\ \hline
Core \%w/w & - & - & - & 84.5 & - & - & - & - & - & 5.6 & 0.3 & 0.6 & 9.0 \\ \hline
\end{tabularx}%
\caption{Elemental composition of the Earth's crust, mantle, and core, in terms of the rounded percentage by weight of each element. We model the Earth's core as a sphere of radius r = 3480 km, enveloped by the mantle extending to r = 6346 km, and then the Earth's crust extending to the Earth's surface at $r = R_\oplus = 6371$ km \cite{WANG2018460, doi:10.1073/pnas.77.12.6973, MCDONOUGH2003547}. This table is reproduced from \cite{acevedo2023oldrocksnewlimits}.}
\label{tab:elements}
\end{table*}

\section{Arrival time distributions for varying composite entry angles}
\label{app:times}

Below we plot constituent arrival times, as in Figure \ref{fig:timingstats}, varying the entry angle $\theta_e$, to show how entry angle affects the distribution of arrival times, for different composite model parameters.

\begin{figure*}[h!]
    \centering
    \includegraphics[width=\textwidth]{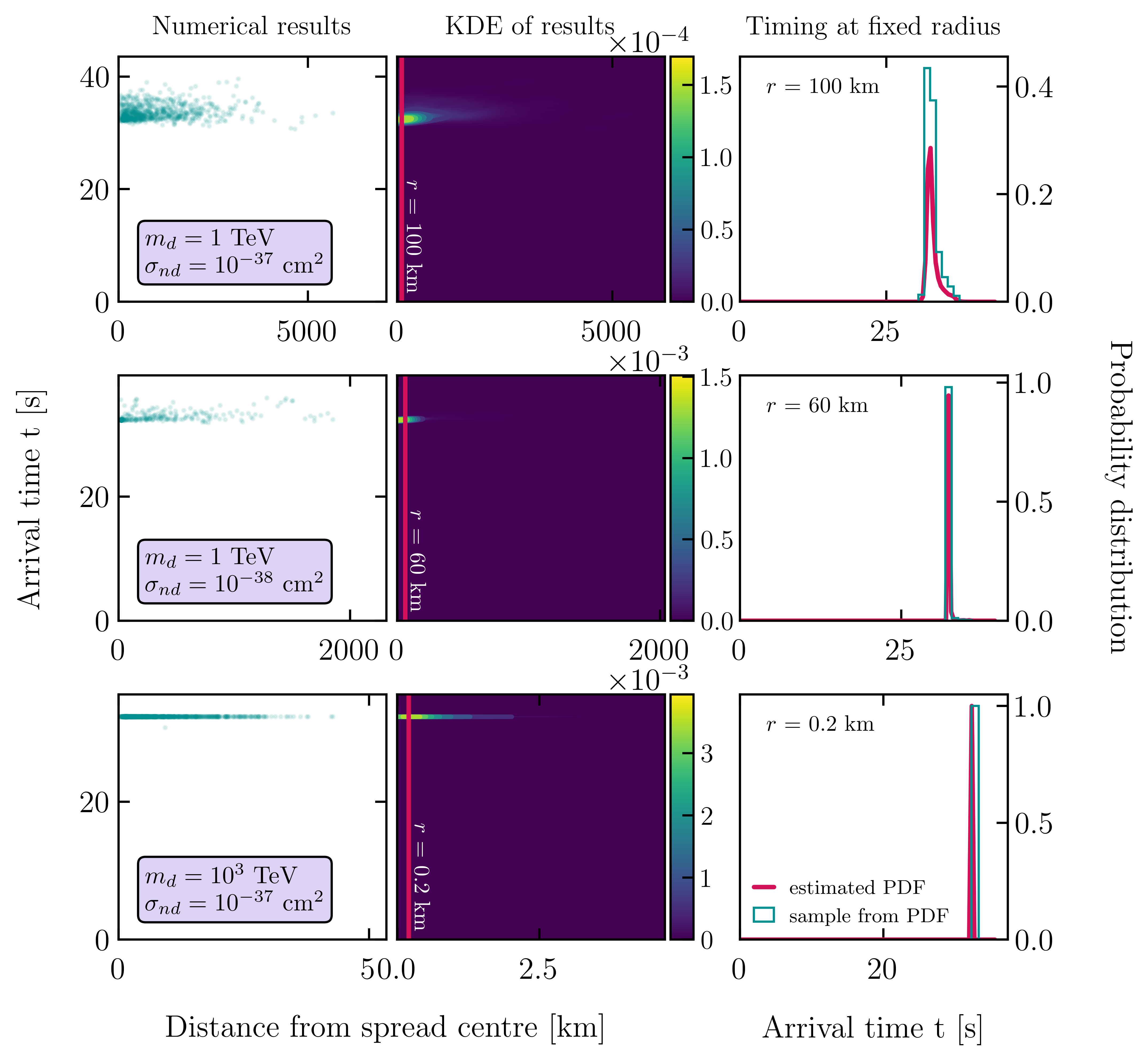}
    \caption{Same as Figure \ref{fig:timingstats}, but with entry angle $\theta_e = 40$ degrees.}
\end{figure*}

\begin{figure*}[h!]
    \centering
    \includegraphics[width=\textwidth]{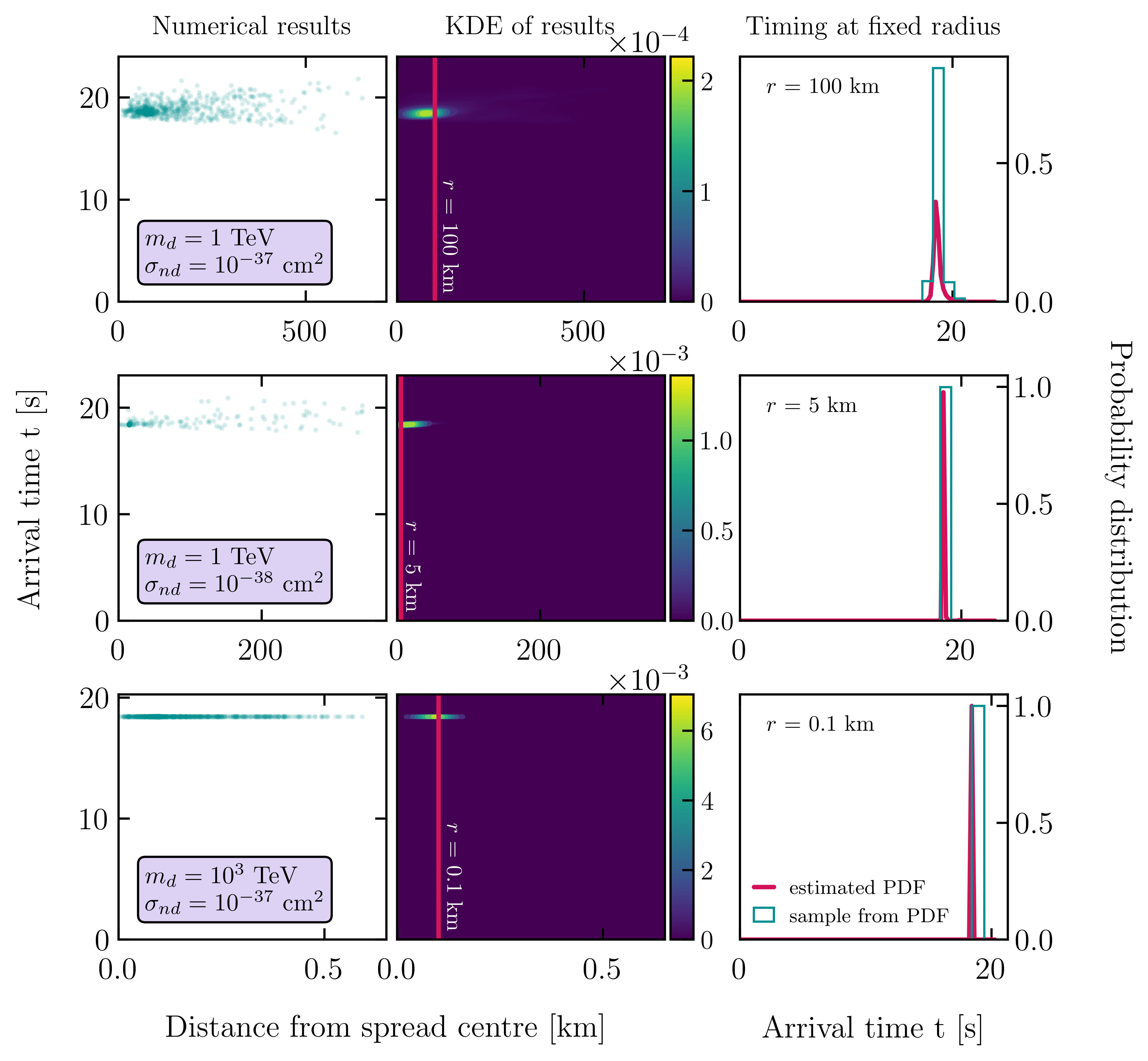}
    \caption{Same as Figure \ref{fig:timingstats}, but with entry angle $\theta_e = 60$ degrees.}
\end{figure*}

\end{document}